\documentclass{iau307}
\usepackage{graphicx}
\usepackage{natbib}
\usepackage{url}
\usepackage{dtklogos}
\bibpunct{(}{)}{;}{a}{}{,}

\title[Massive Star Asteroseismology in Action] 
{Massive Star Asteroseismology in Action}

\author[Conny Aerts]   
{Conny Aerts$^{1,2}$}

\affiliation{$^1$Institute of Astronomy, KU\,Leuven, Celestijnenlaan 200\,D,
  3001 Leuven, Belgium \\ email: {\tt Conny.Aerts@ster.kuleuven.be} \\[\affilskip]
$^2$Department of Astrophysics/IMAPP, Radboud University Nijmegen, 6500 GL
Nijmegen,  the Netherlands}

\pubyear{2014}
\volume{307} 
\pagerange{}
\jname{New windows on massive stars: asteroseismology, interferometry, and 
spectropolarimetry}
\editors{G. Meynet, C. Georgy,  J.H.  Groh \& Ph. Stee, eds.}

\begin{document}

\maketitle

\begin{abstract}
  After highlighting the principle and power of asteroseismology for stellar
  physics, we briefly emphasize some recent progress in this research for various
  types of stars. We give an overview of high-precision high duty-cycle space
  photometry of OB-type stars.  Further, we update the overview of seismic
  estimates of stellar parameters of OB dwarfs, with specific emphasis on
  convective core overshoot. We discuss connections
  between pulsational, rotational, and magnetic variability of massive stars and
  end with future prospects for asteroseismology of evolved OB stars.  
\keywords{ asteroseismology, stars:
    oscillations (including pulsations), stars: interiors, stars: evolution,
    stars: rotation, methods: data analysis, methods: statistical, magnetic
    fields, waves, turbulence}
\end{abstract}

\firstsection 
\section{Progress in Asteroseismology}

\begin{figure}
\begin{center}
\includegraphics[width=0.66\textwidth]{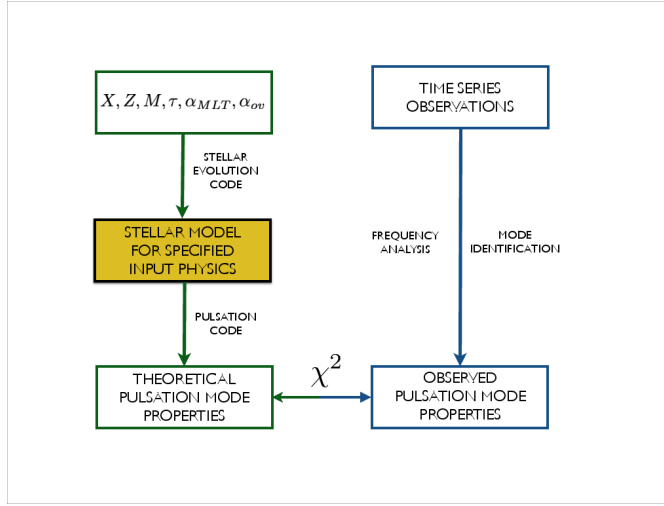} 
\caption{The principle of asteroseismic modelling in a snapshot, where the
  dependency on the chosen input physics of the equilibrium models is
  highlighted in shaded grey. Figure courtesy of Dr.\ Katrijn Cl\'emer.}
\label{fig1}
\end{center}
\end{figure}

Asteroseismology is a fairly new and powerful way of studying stellar physics,
including important excursions to exoplanetary science and the study of galactic
structure and evolution. Even in the context of stellar interiors alone, there
are many aspects to this research field \citep[e.g.,][for an extensive
monograph]{Aerts2010a}. A single snapshot picture capturing the basic idea of
seismic modelling is provided in Fig.\,\ref{fig1}, where $X,Z,M,\tau$ are the
initial hydrogen fraction and metallicity, the mass, and the age, respectively.
In this sketch, we assumed the model computations to be based on the simplistic
(but practical!)  one-dimensional time-independent mixing-length theory of
convection described by the two free parameters $\alpha_{\rm MLT}$ and
$\alpha_{\rm ov}$, which are the mixing length and convective overshoot
parameters, both expressed in units of the local pressure scale height.  For
core-hydrogen burning massive stars, the choice of $\alpha_{\rm MLT}$ is of
limited importance as long as it is within reasonable limits (e.g., between 1.5
and 2.0), while the poorly known amount of convective overshooting expressed by
$\alpha_{\rm ov}$ plays a pivotal role in their structure, in all of the
evolutionary phases. This $\alpha_{\rm ov}$ is thus a key parameter to estimate
and asteroseismology is a relatively new way to do so, as we will discuss
further on.

With the high-precision space photometry from the CoRoT and {\it Kepler\/}
missions at hand for thousands of FGK-type stars, priority was given to the
determination of basic seismic observables and stellar parameters of such stars.
The most popular observables are the so-called large and small frequency
separations, which are the inverse of twice the sound travel time between the
centre and the surface of the star and a measure for the sound-speed gradient in
the stellar interior, respectively.  The frequency of maximum power is also an
important basic observable.  Stochastically-excited modes are easy to identify
as they form specific ridges in so-called \'echelle diagrams, in which the
detected frequencies are reduced modulo the large frequency separation.
Moreover, the seismic diagnostics are easily interpretable in terms of the
fundamental parameters of the stars by means of scaling relations
\citep[e.g.,][]{Kallinger2010,Huber2011,Hekker2011}.  The scheme represented in
Fig.\,\ref{fig1} hence delivers high-precision stellar parameters that are
typically an order of magnitude more precise than those derived from classical
ground-based snapshot spectroscopy or multicolour photometry
\citep[e.g.,][]{Chaplin2014}, even taking into account uncertainties in the
input physics. The seismic parameters derived from scaling relations are
particularly welcomed for host stars of exoplanets
\citep[e.g.,][]{Gilliland2011,Huber2013,Chaplin2013,VanEylen2014,Lebreton2014}
and for galactic clusters and population studies
\citep[e.g.,][]{Corsaro2012,Miglio2013,Stello2013,Casagrande2014}.

The discovery of dipole mixed modes in low-mass
evolved stars \citep{Beck2011} allowed to go beyond the use of simple scaling
relations and led to the derivation of their nuclear burning phase and hence
evolutionary stage \citep{Bedding2011}. Such information is not accessible from
classical data because core-helium burning and hydrogen-shell burning red giants
have the same surface properties. Moreover, after two years of monitoring with
the {\it Kepler\/} satellite, the detected rotational splitting of dipole mixed
modes led to the derivation of their interior rotational properties, with
core rotation typically only a factor 5 to 20 faster than envelope rotation
\citep{Beck2012,Mosser2012,Deheuvels2014}, pointing to major shortcomings in
our understanding of the angular momentum distribution inside stars
(Eggenberger, these proceedings). 

Many remarkable seismic studies of other types of stars are not mentioned here,
due to space constraints and for that reason, we also limit the rest of the paper to
massive stars. Their variability is far more diverse than that of low-mass stars
due to phenomena like fast rotation, mass loss, close binarity, magnetism,
etc. Hence, the case of massive OB star asteroseismology is challenging. A
specific challenge is connected with the nature of the oscillations, i.e., the
majority of the detected oscillation modes in massive stars are self-driven by a
heat mechanism, which operates along with a yet unknown mode selection mechanism
\citep[e.g.,][]{Aerts2010a}.  This creates sparse low-order mode frequency
spectra roughly in the range of 30 to 200\,$\mu$Hz and/or dense high-order
gravity-mode spectra at low frequencies, typically below $20\,\mu$Hz. Moreover,
the rotation of most massive stars is such that their rotationally split
pulsation mode multiplets get merged in frequency spectra, preventing pattern
recognition as for high-order pressure modes in low-mass stars from \'echelle
diagrams. In addition to these inherent difficulties based on the physics of the
pulsations of such stars, OB-type targets have been avoided as much as possible
in the fields of the CoRoT and {\it Kepler\/} exoplanet programmes, to maximise
the efficiency of the search for transiting planets. Despite this limitation,
progress in the asteroseismology of OB stars the past decade is large and is our
point of focus in the remainder of this paper.

\section{Asteroseismic Data of OB Stars in the Space Age}

OB stars have been part of the space photometry revolution since a decade now,
with observations from the MOST, CoRoT, and {\it Kepler\/} satellites. We review
the observational studies based on space photometry and refer to the mentioned
papers for the exquisite light curves and the accompanying Fourier spectra of
the oscillation frequencies.

\subsection{Core-Hydrogen Burning Massive Stars}

MOST was a pioneering mission for massive stars.  Although it could only monitor
stars for typically up to 6 weeks, resulting in limited frequency precision, and
with relatively low brightness precision (typically 0.1\,mmag), it allowed to detect at
least twice as many modes than known from the ground for selected $\beta\,$Cep
stars \citep{Aerts2006a,Handler2009}, to detect pulsations in Be stars
\citep{Saio2007}, and to discover new Slowly Pulsating B stars
\citep{Aerts2006b,Cameron2008,Gruber2012}.

The asteroseismology programme of the CoRoT mission consisted of long runs of
five months (frequency precision $\sim 0.1\mu$Hz) as well as short runs of a few
weeks, dedicated to bright stars (visual magnitude between 6 and 9) resulting in
amplitude precision of roughly $10\mu$mag.  This programme included detailed
studies of carefully selected B stars and showed that the majority of them is
pulsating in numerous modes, some of which with clear period spacings
\citep{Degroote2010b,Papics2012} while others have less structured dense
frequency spectra \citep{Degroote2009a,Thoul2013}.  Some of the B stars turned
out to be rotational variables without pulsations
\citep{Degroote2011,Papics2011}.  One short run of CoRoT was dedicated to the
monitoring of O stars during three weeks. These data showed diversity in the
variability patterns with one $\beta\,$Cep-type pulsator \citep{Briquet2011},
one primary of a long-period binary with stochastic modes \citep{Degroote2010a},
three stars with power excess at low frequency of unknown origin
\citep{Blomme2011} and one complex O-type binary, also known as Plaskett's star
\citep{Mahy2011}. In addition, the CoRoT asteroseismology programme contained a
number of Be stars
\citep{Neiner2009,Juan2009,Diago2009,Huat2009,Desmet2010,Emilio2010,Neiner2012b}. These
Be stars
also revealed quite diverse variability patterns, including an outburst measured
in real time for the B0.5IVe star HD\,49330 due to the beating of non-radial
pulsation modes, stochastically excited gravito-inertial modes, rotational
modulation, and accretion phenomena in a close binary.

Numerous B dwarfs were also found in the CoRoT exoplanet programme
\citep{Degroote2009b}, after variability classification by means of multivariate
Gaussian mixtures \citep{Debosscher2009,Sarro2009} supported by ground-based
follow-up spectroscopy \citep{Sarro2013}. Similar studies have been done for the
{\it Kepler\/} field \citep{Debosscher2011}, along with manual searches for B
pulsators \citep{Balona2011}, resulting in the detection of variability with
similar precision in amplitude than with CoRoT but with a frequency resolution $\sim
0.01\mu$Hz following the four years of monitoring. It is thanks to this
excellent frequency precision that the detection of rotational splitting in
slow rotators became a reality \citep{Papics2014}, with major
progress and future potential in the seismic modelling of massive stars, as outlined
below for two recently discovered pulsators.

Despite these observational studies for hundreds of OB dwarfs, only 16 of them
could be modelled according to the scheme in Fig.\,\ref{fig1} was achieved, {\it the\/}
major obstacle being the lack of identification of the pulsation modes.
However, several promising new cases based on four years of {\it Kepler\/} data
are currently emerging.  As a side remark, the detection of the numerous
pulsation modes at low frequency in so many OB dwarfs, which was not possible
from ground-based data, prompted the need for additional excitation mechanisms
because the classical heat mechanism is not able to explain all the detected
oscillations. One way to solve this is to increase the opacity, either globally
in the star or in the excitation layer \citep{Salmon2012,Walczak2013}.  A
promising new excitation mechanism concerns stochastic gravity modes triggered
by core convection \citep{Belkacem2010,Samadi2010,Saio2011, Shiode2013},
although it remains unclear what their amplitudes are at the stellar surface.

\subsection{Evolved Massive Stars}

The case of asteroseismology of evolved OB stars is far more challenging, but
accordingly more interesting, with large future potential. The challenge is not only due
to the longer oscillation periods (which can reach up to several months) but
also because the oscillations are influenced by various poorly understood physical
processes which take place in the outer atmosphere and wind of such objects,
affecting the boundary conditions that are of importance for the theoretical
predictions needed to perform the scheme in Fig.\,\ref{fig1}.  

Seismic data of evolved OB stars are scarce, even in the space age of
asteroseismology.  Also on the front of supergiant photometric data from space,
MOST delivered the first interesting case with a 37\,d light curve of the
B2Ib/II star HD\,163899. This revealed the simultaneous excitation of pressure
and gravity modes with frequencies below $30\mu$Hz and amplitudes of a few mmag
\citep{Saio2006}.  A MOST campaign of 28\,d dedicated to the B8Ia supergiant
Rigel was too short to uncover all the gravity modes found from six years of
spectroscopic monitoring with frequencies as low as $0.2\mu$Hz and up to
$12\mu$Hz \citep{Moravveji2012}.  Of a completely different nature is the 137\,d
CoRoT light curve of the B6I star HD\,50064, which, combined with spectroscopy,
led to the conclusion that periodic mass-loss episodes due to a variable
large-amplitude mode seem to occur \citep{Aerts2010b}.  Finally, a 26\,d CoRoT light
curve of the B8Ib star HD\,46769 led to the detection of low-amplitude
rotational modulation with a frequency of $2.4\mu$Hz and an amplitude of some
100$\mu$mag rather than pulsations \citep{Aerts2013a}.

The only evolved massive star observed by the {\it Kepler\/} mission is the bright
eclipsing binary V380\,Cyg, which required a customized mask
\citep{Tkachenko2012}. Its {\it Kepler\/} light curve, along with extensive time-resolved
spectroscopy led to the detection of rotational modulation along with
low-amplitude stochastic variability at low frequency in the primary
\citep{Tkachenko2014b}. This variability is of similar nature than the patterns
detected in the three hottest CoRoT O dwarfs \citep{Blomme2011} and is
compatible with theoretical predictions for stochastic gravity waves excited by
the convective core. 

The detailed theoretical interpretation of the periodic oscillations in
supergiants remains to be done.  While \citet{Godart2009} showed that mass loss
and large overshooting reduce the extent of an otherwise occurring intermediate
convective zone, preventing the excitation of gravity modes due to a heat
mechanism acting in the metal opacity bump, \citet{Daszynska2013} do find such
modes irrespective of an intermediate convection zone or not.  Theoretical work
on the improvement of stellar structure models of evolved massive stars will
benefit from the seismic and spectroscopic data, following the scheme in
Fig.\,\ref{fig1}.

\section{Seismic Modelling of OB Dwarfs\label{modelling}}
\begin{figure}
\begin{center}
\rotatebox{270}{\resizebox{9cm}{!}{\includegraphics{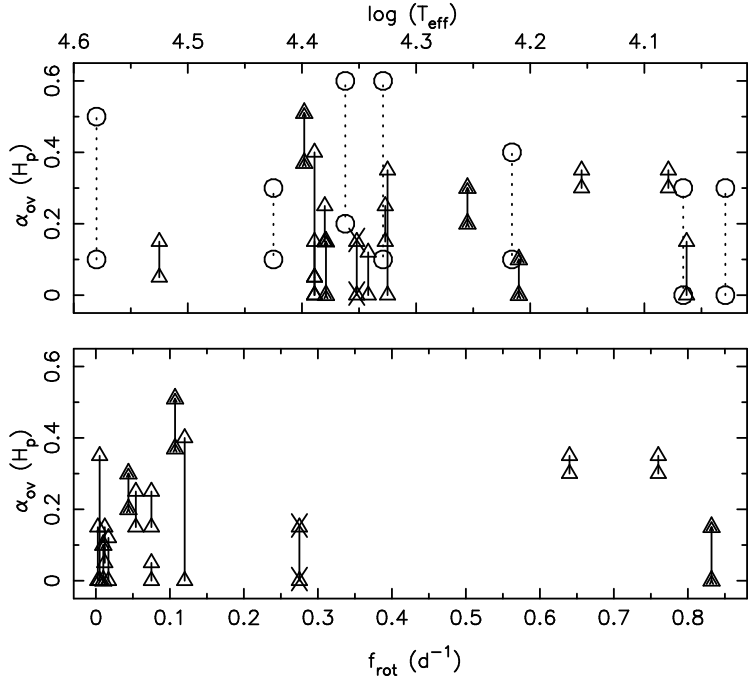}}}
\caption{Core overshoot parameters from asteroseismology of OB stars as a
  function of the rotational frequency (lower panel) and effective temperature
  (upper panel), where single stars are indicated by open triangles and
  spectroscopic binaries by filled triangles; the magnetic pulsator is indicated
  with an additional cross. The estimates for the primaries of
  eclipsing binaries \citep{Claret2007} are indicated in the upper panel by
  circles connected with dotted lines.}
\label{fig2}
\end{center}
\end{figure}

Unlike solar-like stars and red giants, where ensemble asteroseismology can be
achieved from scaling relations, the seismic modelling of massive stars requires
a star-by-star treatment of the scheme in Fig.\,\ref{fig1} and is thus immensely
work-intensive.  \citet{Aerts2013b} provided a compilation of the seismically
derived stellar parameters according to Fig.\,\ref{fig1}, including the core
overshoot value assuming the Schwarzschild criterion of convection and a fully
mixed overshoot region, for 8 single and 3 binary non-emission OB dwarfs, among
which one magnetic pulsator. The majority of these studies was based on
multisite campaigns. In addition, seismic modelling was achieved for
the late Be stars HD\,181231 and HD\,175869 \citep{Neiner2012a}.  New studies
since then were done for the $\beta\,$Cep stars $\gamma\,$Peg
\citep{Walczak2013} and $\sigma\,$Sco \citep{Tkachenko2014b}, as well as for the
B8.3V star KIC\,10526294 \citep{Papics2014}. The latter star is a young slowly
rotating ($P_{\rm rot}=190$\,d) B star with 19 rotationally split and quasi-equally
spaced dipole gravity modes detected in the 4-year {\it Kepler\/} light
curve. Frequency inversion of its triplets leads to counter-rotation in its envelope
(Triana et al., submitted).  This star, along with the A-type pulsator
KIC\,11145123 whose envelope rotates slightly faster than its core, as revealed
by rotationally split triplets and quintuplets \citep{Kurtz2014}, will
undoubtedly give rise to future improvement of the input physics for stars of
intermediate mass, because the current models cannot
explain the observed pulsational properties.

A major finding from asteroseismology and of relevance for stellar evolution of
massive stars is the need for extra mixing, enlarging the core masses of OB
dwarfs, already from their early life. This was also hinted at long ago from
modelling of the turn-off point of clusters, but this method has limited
predictive power due to observational biases.  Isochrone fitting of eclipsing
binaries is a more powerful method to pinpoint $\alpha_{\rm ov}$
\citep{Claret2007,Torres2014} independently from asteroseismology. In
Fig.\,\ref{fig2} we provide a compilation of the $\alpha_{\rm ov}$ values
derived from asteroseismology for 16 OB dwarfs and from isochrone fitting for 7
unevolved eclipsing binaries. Only 9 of the 23 stars are compatible with the
absence of core overshoot. Other than that, no obvious relation emerges. The
same conclusion is found when comparing $\alpha_{\rm ov}$ with the mass.


\section{Spin-offs of Massive Star Asteroseismology}

\subsection{Macroturbulence}

The wings of the profiles of metal lines of a large fraction of OB stars cannot
be explained by 1-dimensional atmosphere models, unless a macroturbulent
velocity field is added (S\'{\i}mon-D\'{\i}az, these proceedings). Independently
of whether one uses an isotropic or a radial-tangential Gaussian
\citep{Gray2005} to describe this macroturbulence, supersonic speeds are often
needed to bring the data in agreement with the model predictions, particularly
for O stars and supergiants.

\citet{Aerts2009} suggested that the collective pulsational velocity broadening
due to gravity modes could be a viable physical explanation for macroturbulence
in massive stars. However, while B dwarfs are known to undergo pulsations with
accompanying periodic line-profile variations
\citep[e.g.,][]{DeCatAerts2002,AertsDeCat2003}, pulsations with the required
frequencies and velocity amplitudes have yet to be firmly established for large
samples of O stars and B supergiants.  Indeed, the samples of OB stars observed
by \citet{SimonDiaz2014} and \citet{Markova2014} reveal a diversity of
time-variability in terms of line broadening, with the dwarfs having clear
line-profile variations but the evolved stars not necessarily so while their
line wings are strongly broadened. It therefore seems that convective velocities
in the outer envelope
are a more plausible explanation than pulsations for evolved stars
(S\'{\i}mon-D\'{\i}az, these proceedings).


\subsection{Multivariate Statistical Analysis of Oscillations, Rotation, and
  Magnetic Field} 

In the spirit of this symposium of considering an integrated approach,
\citet{Aerts2014} made a multivariate study by combining spectropolarimetric,
asteroseismic and spectroscopic observables derived from high-precision data for
64 galactic dwarfs. This 10-dimensional data set is complete in $\log T_{\rm
  eff}$, $\log g$, and $v\sin i$ and is more than 94\% complete in the
rotational frequency $f_{\rm rot}$ and the magnetic field strength. An estimate
of the nitrogen abundance is available for 59\% of the targets while 32\% of the
stars have gravity-mode oscillation frequencies and 32\% have pressure-mode
oscillation frequencies. The stars without oscillations all have
tight upper limits on the amplitudes.

One of the major aims of this study was to search for correlations between the
nitrogen abundance and the other nine observables, from multivariate data
analysis.  This was achieved by applying a statistical technique called {\it
  multiple imputation\/} for the missing values, followed by linear regression
from both backward and forward selection.

No significant correlation was found between the rotational observables $v\sin
i$ or $f_{\rm rot}$ and the nitrogen abundance. The latter did correlate with
$\log T_{\rm eff}$ and with the dominant acoustic oscillation frequency of the stars.
Also a correlation between the dominant gravity-mode frequency and the magnetic
field strength was found, but none of these two observables beared any relation
with the nitrogen abundance.

While the sample may be prone to biases in the spectroscopic observables (cf.\
Przybilla, these proceedings), the study showed that rotation cannot be the only
physical process to cause mixing in stellar interiors of galactic dwarfs and
that magnetic fields are not a good alternative explanation.  The role of other
phenomena causing chemical mixing, such as oscillation modes or internal gravity
waves (Mathis, these proceedings) could be a viable alternative or additional
explanation.



\section{Conclusions and Future Prospects}

The past five years, photometric asteroseismic data improved in precision from
mmag to $\mu$mag and in duty cycle from tens of percents to more than
90\%. Moreover, data became available for thousands of stars in almost all
evolutionary stages, instead of being limited to a handful of solar-like stars,
young B stars, subdwarfs and white dwarfs.
As a consequence, asteroseismology opened a new window for stellar physics in 
such a way that
stellar structure and evolution studies are currently observationally driven.

The discovery of gravity-mode oscillations in thousands of CoRoT and {\it
  Kepler\/} targets allows to probe the physics in and near stellar cores in
both red giants (Eggenberger, these proceedings) and in massive stars. A major
finding is that the internal stellar rotation, and along with it the angular
momentum distribution inside stars, is not at all what current theories predict
it should be. While the deviation between theory and observations in terms of
the interior rotation profile in the A-type star KIC\,11145123 and in the B-type
star KIC\,10526294 remains modest and can qualitatively be explained by the act
of internal gravity waves (Mathis, these proceedings), the discrepancy for red
giants is two orders of magnitude and requires at least one hitherto omitted
strong coupling mechanism between the stellar core and the envelope (see also
Maeder, these proceedings).

While asteroseismology of massive stars has been in action since more than a
decade now, after the first seismic probing of the core overshooting
\citep{Aerts2003a}, the number of stars with detailed tuning of the internal
physics is far too low and, moreover, restricted to dwarfs.  This is first of
all due to the limited number of such stars in the fields of CoRoT and {\it
  Kepler\/} (some tens compared to many thousands of giants), but also has to do
with the lack of mode identification for the majority of OB-type stars observed
in white-light space photometry. The problem of mode identification can only be
overcome from extensive data sets of multicolour photometry or high-resolution
high-precision spectroscopy of bright stars \citep[][Chapter 6]{Aerts2010b},
from the detection of period spacings of high-order gravity modes,
\citep[e.g.,][]{Degroote2010a,Papics2012}, from the occurrence of rotationally
split multiplets \citep[e.g.,][]{Aerts2003a,Pamyatnykh2004,Briquet2007b}, or,
ideally, a combination of the latter two cases \citep{Papics2014}.  The
detection of rotationally split multiplets of high-order gravity modes was only
achieved for two dwarfs with a convective core so far, and only after analysing
four years of {\it Kepler\/} data \citep{Kurtz2014,Papics2014}. These two stars
will certainly be modelled in more detail in the coming months, with attention
to improvement of the input physics of models, because the current theory cannot
explain the details of the observed seismic behaviour (Moravveji, these proceedings).

A new opportunity in the coming months and years, are several hundreds of
massive OB stars to be observed by the {\it Kepler\/} 2-wheel mission, a.k.a.\
K2 \citep{Howell2014}.  It remains to be seen how many of those OB-type targets
turn out to be pulsators with rotational splitting, given the rather limited
time base of three months of continuous monitoring, but even a low percentage of
the hundreds of submitted OB stars will open new avenues for massive star
seismology including numerous evolved OB stars.

Another near-future avenue for improved seismic modelling is the addition of an
accurate distance, and by implication a model-independent luminosity and radius
estimate, from the ESA cornerstone mission Gaia (launched 19 December 2013 and
currently in its commissioning phase). A direct accurate angular diameter
measurement for bright stars from interferometry (van Belle, these proceedings),
combined with an accurate parallax determination, would have similar
capacity. The combination of asteroseismic data and a model-independent radius
estimate implies a serious reduction in the uncertainties of the physical
quantities of the stars from seismic inference, as outlined in detail by
\citet{Cunha2007} and \citet{Huber2012} for stars with stochastically-excited
oscillations. Given the current lack of accurate distances for massive stars,
and of good calibrators for a high-precision interferometric radius
determination of pulsating OB stars, progress in this area is to be expected in 
the next years.

  Finally, the recently approved ESA M3 mission PLATO \citep{Rauer2014}, to be
  launched in 2024, will measure numerous OB-type stars in its very wide
  Field-of-View (2250\,deg$^2$). This will include high-cadence (2.5\,s)
  two-colour measurements of stars with visual brightness between 4 and 8 to be
  obtained by its two bright-star telescopes with a 95\% duty cycle for a
  minimum of 2 years and, possibly, fainter OB stars in white light with a cadence
  of 32\,s observable with all of the 32 normal telescopes, during the
  step-and-stare phase of the mission.


\bibliographystyle{iau307}
\bibliography{IAUS307_Aerts_astroph}

\end{document}